\documentstyle[12pt,epsf]{article}

\global\arraycolsep=1pt
\def\be{\begin{equation}}
\def\ee{\end{equation}}
\def\bea{\begin{eqnarray}}
\def\eea{\end{eqnarray}}
\def\w{\omega}
\def\gm{\gamma}
\def\hf{{1 \over 2}}
\begin{document}
\begin{flushright}
OU-HET 293 \\ DTP-98-25 \\ {\bf math-ph/9805007} 
\end{flushright}
\vskip 0.8 cm

\begin{Large}
\centerline{{\bf Integrable Top Equations associated}}
\vskip 0.1cm
\centerline{{\bf with Projective Geometry over $Z_2$}}
\end{Large}
\vskip 1.0cm

\centerline{\ \ David B. Fairlie$^\dag$ and Tatsuya Ueno$^\ddag$}
\vskip 0.5cm

\centerline{{$^\dag$}Department of Mathematical Sciences,}
\vskip 5pt
\centerline{University of Durham,}
\vskip 5pt
\centerline{Durham, DH1 3LE, England.}
\vskip 5pt
\centerline{{\sl david.fairlie@durham.ac.uk}} 
\vskip 10pt

\centerline{{$^\ddag$}Department of Physics, Graduate School of 
Science,}
\vskip 5pt
\centerline{Osaka University,}
\vskip 5pt
\centerline{Toyonaka, Osaka 560-0043, Japan.}
\vskip 5pt
\centerline{\quad{\sl ueno@graviton.phys.sci.osaka-u.ac.jp}} 
\centerline{\quad{\sl tatsuya@yukawa.kyoto-u.ac.jp}} 
\vskip 2.0cm

\begin{center}
{\bf ABSTRACT}
\end{center}
We give a series of integrable top equations associated with
the projective geometry over $Z_2$ as a $(2^n-1)$-dimensional
generalisation of the 3D Euler top equations.
The general solution of the $(2^n-1)$D top is shown to be given
by an integration over a Riemann surface with genus 
$(2^{n-1}-1)^2$.
\newpage

\section{${\bf (2^{{\bf \it n}}-1)}$D top equations}%
Recently we discovered an apparently new integrable set of 
evolution equations in seven dimensions, which are an analogue of 
the well-known 3D Euler top \cite{dbf}\cite{tu}. 
The 7D top arises from the dimensional reduction of the 8D 
$Spin(7)$ invariant self-dual Yang-Mills (SDYM) equations in 
\cite{cdfn}, just as the 3D top comes from the reduction of the 
4D SDYM to differential equations depending only upon one variable.
The integrability of the 3D top is ensured by the existence of the Lax
formulation of the 4D SDYM \cite{ac}, while there is no such
first order structure behind the 8D SDYM \cite{rsw}.
Nevertheless the 7D top has been shown to have sufficient conserved 
quantities to permit full integrability \cite{dbf} and its general 
solution is given by a non-hyperelliptic differential equation 
corresponding to a Riemann surface with genus 9 \cite{tu}.
\par

The derivation of the top equations from the SDYM shows their 
connection with the existence of the division algebras, 
the 3D system arising from the quaternionic algebra, the 7D one 
from the octonions, which seems to suggest that no further 
integrable top system in more than seven dimensions should exist.
In this note, however, we demonstrate that a generalisation of our 
previous results to general $2^n-1$ dimensions is possible and is 
associated rather with the $n$-dimensional projective space over 
$Z_2$. 
\par

We take the projective space $Z_2P_{n-1}$ with homogeneous 
coordinates $(z_0,z_1,\cdots,z_{n-1})$, where $z_i$ is either 0,1 and 
calculations are performed in arithmetic mod 2. 
The space $Z_2P_{n-1}$ consists of a finite number of points $e_i$ 
$(i=1,\cdots,2^n-1)$ with the multiplication operation $e_ie_j$ defined 
by the sum of their associated coordinates. 
\par

For the 3D ($n=2$) case, we have three points,
\be
e_1 = (0,1) \ , \qquad e_2 = (1,0) \ , \qquad e_3 = (1,1) \ ,
\ee
with the multiplication rule,
\be
e_i e_j = \varepsilon_{ijk}^2 \, e_k \ , 
\ee
where $\varepsilon_{ijk}$ is the structure constant of 
the $su(2)$ (quaternion) algebra. 
Using this structure constant, we obtain the 3D Euler top 
equations with variables $(\w_1(t),\w_2(t),\w_3(t))$,
\be
{d \over dt}\w_i = \hf \varepsilon_{ijk}^2 \, \w_j \, \w_k \ .
\label{euler}
\ee
\par

In the 7D ($n=3$) case, we have seven points,
\bea
&&e_1 = (0,0,1) \ , \ e_2 = (0,1,0) \ , \ e_3 = (1,0,0) \ , \ 
e_4 = (1,1,1) \ , \nonumber \\
&&e_5 = (1,1,0) \ , \ e_6 = (1,0,1) \ , \ e_7 = (0,1,1) \ ,
\eea
with the relation
\be
e_i e_j = c_{ijk}^2 \, e_k               \label{oct2}
\ee
where $c_{ijk}$ is equal to a realization of the totally 
anti-symmetric structure constant appearing in the Cayley
(octonion) algebra,
\be
c_{127}=c_{631}=c_{541}=c_{532}=c_{246}=c_{734}
=c_{567}=1 \ .  \qquad
(\rm{others \ zero})
\ee
The relation (\ref{oct2}) can be read off from the diagram
in Fig.1, the seven-point plane with 7 points and 7 lines;
3 points lie on each line and 3 lines pass through each point.
\begin{figure}[b]
\epsfxsize= 45 mm
\begin{center}
\leavevmode
\epsfbox{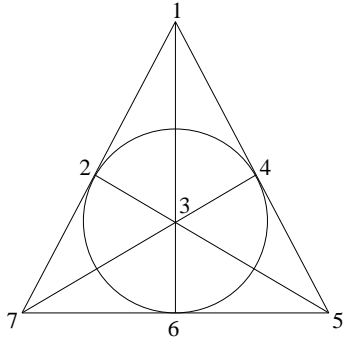}
\end{center}
\caption{7 point plane.}
\end{figure}
Replacing $\varepsilon_{ijk}$ in (\ref{euler}) by the constant
$c_{ijk}$, we obtain the set of seven equations for a 7D top 
\cite{dbf}\cite{tu},
\bea
&&
{d \over dt}\w_1= \w_2\w_7+\w_6\w_3+\w_5\w_4 \ ,
\quad
{d \over dt}\w_2=\w_7\w_1+\w_5\w_3+\w_4\w_6 \ ,
\nonumber\\
&&{d \over dt}\w_3=\w_1\w_6+\w_2\w_5+\w_4\w_7 \ ,
\quad
{d \over dt}\w_4=\w_1\w_5+\w_6\w_2+\w_7\w_3 \ ,
\label{euler7}\\
&&{d \over dt}\w_5=\w_4\w_1+\w_3\w_2+\w_6\w_7 \ ,
\quad
{d \over dt}\w_6 =\w_3\w_1+\w_2\w_4+\w_7\w_5 \ ,
\nonumber\\
&&{d \over dt}\w_7=\w_1\w_2+\w_3\w_4+\w_5\w_6 \ .
\nonumber
\eea
\par

In a similar fashion to the above 3D and 7D cases, we can obtain
$2^n-1$ equations for a $(2^n-1)$D top. 
The structure of the higher-dimensional tops can be understood from 
the $(2^n-1)$-point hyperplane diagram, which is an extension of the 
seven-point plane and consists of $2^n-1$ points and $2^n-1$ 
$(2^{n-1}-1)$-point hyperplanes, where $2^{n-1}-1$ points lie on each 
$(2^{n-1}-1)$-point plane and $2^{n-1}-1$ $(2^{n-1}-1)$-point 
hyperplanes pass through each point.
\par

For example, in the 15D ($n=4$) case with an appropriate labelling
of 15 points in $Z_2P_3$, we have a 15-point tetrahedral space 
containing the following 15 7-point planes assigned by 7 points in 
them,
\bea
&&(1,2,3,4,5,6,7) \ , \ \ \ \ \ \, (1,2,8,11,10,9,7) \ , \ \ \ \ \
(1,3,8,13,12,9,6) \ , \nonumber \\
&&(2,3,8,14,12,10,5) \ , \ (1,2,13,14,15,12,7) \ , \ \ 
(1,3,14,11,10,15,6) \ , \nonumber \\
&&(1,4,8,14,15,9,5) \ , \ \ \, (1,4,13,11,10,12,5) \ , \ \ 
(2,3,11,13,15,9,5) \ , \\
&&(2,4,8,13,15,10,6) \ , \  (2,4,11,14,12,9,6) \ , \ \ \ \,
(3,4,8,11,15,12,7) \ ,  \nonumber \\
&&(3,4,9,10,14,13,7) \ , \ (5,6,8,11,13,14,7) \ , \ \ \ \,
(5,6,9,10,12,15,7) \ ,  \nonumber 
\eea
where the $p$-th element in each of the above 15 brackets
is placed on the point $p$ in Fig.1.
The form of the 15D top equations derived from the 15 
point-hyperplane is 
\bea
&&\dot{\w}_1 = \w_2 \w_7 + \w_3 \w_6 + \w_5 \w_4 + \w_8 \w_9 + 
\w_{10}\w_{11} + \w_{12}\w_{13} + \w_{14}\w_{15} \ ,
\nonumber \\
&&\dot{\w}_2 = \w_1 \w_7 + \w_3 \w_5 + \w_4\w_6 + \w_8 \w_{10} + 
\w_{11}\w_{9} + \w_{12}\w_{14} + \w_{15}\w_{13} \ ,
\nonumber \\
&&\dot{\w}_3 = \w_1 \w_6 + \w_2 \w_5 + \w_7 \w_4 + \w_8 \w_{12} + 
\w_{9}\w_{13} + \w_{10}\w_{14} + \w_{11}\w_{15} \ ,
\nonumber \\
&&\dot{\w}_4 = \w_5 \w_1 + \w_2 \w_6 + \w_7 \w_3 + \w_8 \w_{15} + 
\w_9\w_{14} + \w_{10}\w_{13} + \w_{11}\w_{12} \ ,
\nonumber \\
&&\dot{\w}_5 = \w_1 \w_4 + \w_2 \w_3 + \w_7 \w_6 + \w_8 \w_{14} + 
\w_9\w_{15} + \w_{10}\w_{12} + \w_{11}\w_{13} \ ,
\nonumber \\
&&\dot{\w}_6 = \w_1 \w_3 + \w_2 \w_4 + \w_7 \w_5 + \w_8 \w_{13} + 
\w_9\w_{12} + \w_{10}\w_{15} + \w_{11}\w_{14} \ ,
\nonumber \\
&&\dot{\w}_7 = \w_1 \w_2 + \w_3 \w_4 + \w_6 \w_5 + \w_8 \w_{11} + 
\w_9\w_{10} + \w_{12}\w_{15} + \w_{13}\w_{14} \ ,
\nonumber \\
&&\dot{\w}_8 = \w_1 \w_9 + \w_2 \w_{10} + \w_3 \w_{12} 
+ \w_4 \w_{15}+ \w_5\w_{14} + \w_6\w_{13} + \w_7\w_{11} \ ,
\\
&&\dot{\w}_9 = \w_1 \w_8 + \w_2 \w_{11} + \w_3 \w_{13} 
+ \w_4 \w_{14}+ \w_5\w_{15} + \w_6\w_{12} + \w_7\w_{10} \ ,
\nonumber \\
&&\dot{\w}_{10} = \w_1 \w_{11} + \w_2 \w_8 + \w_3 \w_{14} 
+ \w_4 \w_{13} + \w_5\w_{12} + \w_6 \w_{15} + \w_7 \w_9 \ ,
\nonumber \\
&&\dot{\w}_{11} = \w_1 \w_{10} + \w_2 \w_9 + \w_3 \w_{15} 
+ \w_4 \w_{12} + \w_5\w_{13} + \w_6\w_{14} + \w_7\w_8 \ ,
\nonumber \\
&&\dot{\w}_{12} = \w_1 \w_{13} + \w_2 \w_{14} + \w_3 \w_8 
+ \w_4 \w_{11} + \w_5\w_{10} + \w_6\w_9 + \w_7\w_{15} \ ,
\nonumber \\
&&\dot{\w}_{13} = \w_1 \w_{12} + \w_2 \w_{15} + \w_3 \w_9 
+ \w_4 \w_{10} + \w_5\w_{11} + \w_6\w_8 + \w_7\w_{14} \ ,
\nonumber \\
&&\dot{\w}_{14} = \w_1 \w_{15} + \w_2 \w_{12} + \w_3 \w_{10}
+ \w_4 \w_9 + \w_5\w_8 + \w_6\w_{11} + \w_7\w_{13} \ ,
\nonumber \\
&&\dot{\w}_{15} = \w_1 \w_{14} + \w_2 \w_{13} + \w_3 \w_{11}
+ \w_4 \w_8 + \w_5\w_9 + \w_6\w_{10} + \w_7\w_{12} \ .
\nonumber 
\eea
\par

\section{General solution of the ${\bf (2^{\bf {\it n}}-1)}$D top}%
\subsection{Integrability}
To show the integrability of the $(2^n-1)$D top and its general 
solution, it is convenient to work with a set of $2^n-1$ variables $a_i$,
instead of the $\w_i$'s.
The rule to define the $a_i$'s is to pick up $2^n-1$ sets of
$2^{n-1}$ $\w_i$'s which do not lie on a $(2^{n-1}-1)$-point 
subplane in the $(2^n-1)$-point hyperplane and to assign $a_i$ to the 
sum of all $2^{n-1}$ $\w_i$'s in each of the $2^n-1$ sets.
For example, in the 3D case,
\be
a_1 = \w_2 + \w_3 \ , \qquad a_2 = \w_3 + \w_1 \ ,
\qquad a_3 = \w_1 + \w_2 \ ,
\ee
and in the 7D case,
\bea
&&a_1 = \w_3+\w_4+\w_5+\w_6 \ , \qquad 
a_2 = \w_1+\w_2+\w_5+\w_6 \ ,
\nonumber \\
&&a_3 = \w_1+\w_3+\w_5+\w_7 \ , \qquad 
a_4 = \w_2+\w_4+\w_5+\w_7 \ ,
\label{abc} \\
&&a_5 = \w_2+\w_3+\w_6+\w_7 \ , \qquad 
a_6 = \w_1+\w_4+\w_6+\w_7 \ ,
\nonumber \\
&&a_7 = \w_1+\w_2+\w_3+\w_4 \ .
\nonumber 
\eea
Similarly, 15 variables $a_i$ in the 15D top can be easily read 
off from the 15-point hyperplane defined in (8).
\par

Using the variables $a_i$, the $(2^n-1)$D top equations are 
re-expressed as
\be
\dot{a}_i =  a_i \, (S -  a_i) \ , \ \ 
\qquad S = {1 \over 2^{n-1}} \sum_{j=1}^{2^n-1}a_j \ ,
\label{eqa}
\ee
and the equations of motion for the difference of the $a_i$'s are  
\be
\dot{(a_i - a_k)} = (a_i - a_k) \, (S - a_i - a_k) \ .
\label{eqaa}
\ee
We introduce the quantity $W$ with the constants $\rho_i$ and
$\chi_{ij}$,
\be
W = \sum_i \rho_i \ln{a_i} + \sum_{i < j}\chi_{ij} \ln{(a_i - a_j)}
\ .
\label{W}
\ee
The condition $\dot{W} = 0$ leads us to $(2^n-1)(2^{n-1}-1)$
conserved quantities $N_{ij}$,
\be
N_{ij} = T (a_i - a_j)/a_ia_j \ , \ \ 
T = (\prod_{k=1}^{2^n-1}a_k)^{1 \over 2^{n-1}-1} \ .
\label{nij}
\ee
Although  the $N_{ij}$ are not independent, they are sufficient to
construct a basis of  $2^n-2$ independent conserved quantities, thus 
guaranteeing the integrability of the $(2^n-1)$D top.
Specifically, all the $N_{ij}$ can be expressed in terms of  $N_{1j}$
$(j=2,\cdots,2^n-1)$ through the relation
\be
N_{ij} = N_{1j} - N_{1i} \ , 
\label{nnn}
\ee
which means that any conserved quantities in the system can be 
constructed from these $2^n-2$ quantities $N_{1j}$.
In particular it is possible to define $2^n-1$ polynomial 
conserved quantities $\gm_i$ from $N_{ij}$ as
\be
\gm_i = N_{j_1k_1}N_{j_2k_2} \cdots 
N_{j_{2^{n-1}-1} k_{2^{n-1}-1}} 
= a_i(a_{j_1}-a_{k_1})(a_{j_2}-a_{k_2}) \cdots 
(a_{j_{2^{n-1}-1}}-a_{k_{2^{n-1}-1}}) \ , 
\ee
where $(j_p,k_p)$, $(j_p < k_p, \ p = 1,\cdots,2^{n-1}-1)$ 
lie on the respective $2^{n-1}-1$ lines through the point $i$.
The polynomials $\gm_i$ are of order $2^{n-1}$.
There is, of course, one functional relationship connecting these  
$2^n-1$ expressions.
\par

Summing over the index $i$ of $N_{ij}$ in (\ref{nij}), we see that 
all $a_j$ are expressed in terms of two variables $T$ and $U$,
with the constants $M_j = \sum_{i=1}^{2^n-1}N_{ij}/(2^n-1)$,
\be
a_j^{-1} = M_j T^{-1} + U \ , \qquad 
U = {1 \over 2^n-1}\sum_{i=1}^{2^n-1}a_i^{-1} \ .
\label{atu}
\ee
Note that the variables $T$ and $U$ are symmetric under any
permutation of $a_i$'s.
Substituting the expression of $a_i$'s into the definition of $T$
in (\ref{nij}), we have the following relation between $T$ and $U$,
\be
T^{2^{n-1}} = \prod_{j=1}^{2^n-1}(T \, U + M_j) \ .
\label{tu}
\ee
{}From (\ref{atu}) and (\ref{tu}), we see that all variables are 
expressible in terms of one variable, which demonstrates that the  
system of the $(2^n-1)$D top is integrable. 
The explicit expression for the quadrature whose evaluation solves 
the top is given in the next subsection. 
\par

\subsection{General solution}
The time derivatives of $T$ and $U$ are derived from the 
equations of motion (\ref{eqa}),
\be
\dot{T} = T \, S \ , \qquad \qquad
\dot{U} = - U \, S + 1 \ .
\ee
We introduce a variable $R(t) = T(t)\, U(t)$, whose time
derivative is given as
\be
\dot{R} = T \ .
\label{rt}
\ee
Substituting $R = T\, U$ into (\ref{tu}) and (\ref{atu}), we have
\be
T = (\prod_{j=1}^{2^n-1}(R+M_j))^{1 \over 2^{n-1}} \ ,
\label{trm}
\ee
and 
\be
a_j =\frac{ T}{(R+M_j) }= 
\frac{(\prod_{k=1}^{2^n-1}(R+M_k))^{1 \over 2^{n-1}}}
{(R+M_j)} 
\ .
\ee
Using (\ref{rt}) and (\ref{trm}), we obtain a first-order 
equation for $R(t)$,
\be
\dot{R} = 
(\prod_{j=1}^{2^n-1}(R+M_j))^{1 \over 2^{n-1}} \ ,
\label{Req}
\ee
which is non-hyperelliptic except for the 3D $(n=2)$ case.
The integral associated with this equation can be shown to 
correspond to a Riemann surface with genus $g=(2^{n-1}-1)^2$;
the order $1/2^{n-1}$ in the RHS of (\ref{Req}) means that 
we need $2^{n-1}$ complex surfaces, each of which has 
$2^{n-1}$ cuts since the order of $R$ is $2^n-1$ inside the 
bracket of the RHS.
\par

\section{Further generalisations}%
It would be natural to expect that the examples of this note could be 
further generalised to the discussion of evolution equations for 
$\displaystyle{\frac{k^{n}-1}{k-1}}$ variables, corresponding to tops 
based upon the space $Z_kP_{n-1}$. 
Despite many efforts, we have as yet been unable to demonstrate a set 
of integrable equations for integral $k>2$ except for the case where 
$n=2$ and there are $k+1$ points lying on a line. 
Then one possibility for a set of integrable evolution equations is 
\cite{dbf2}
\be
\frac{d}{ dt}\w_i =\prod_{j\neq i}\w_j \ . \qquad (i=1,\cdots,k+1)
\ee
These equations are reduced to a hyperelliptic differential equation for a 
$g=k-1$ Riemann surface.
It would be surprising if there is no elegant integrable 
generalisation of these or similar evolution equations.  

\vskip 1.5cm
\section*{Acknowledgements}
T. Ueno is supported by the Japan Society for the Promotion of 
Science, No.\,6293.
\newpage

\end{document}